\documentclass[12pt]{iopart}
\usepackage{graphicx}
\begin{document}

\title[]{A Wave Function Describing Superfluidity in a Perfect Crystal}

\author{Hui Zhai\dag \footnote[3]{electronic address: huizhai@castu.tsinghua.edu.cn}
and Yong-Shi Wu\dag\ddag \footnote[4]{electronic address:
wu@physics.utah.edu}}

\address{\dag\ Center for Advanced Study, Tsinghua University,
Beijing, 100084, P. R. China}

\address{\ddag\ Department of Physics, University of Utah, Salt Lake City,
Utah 84112, U.S.A.}

\begin{abstract}
We propose a many-body wave function that exhibits both diagonal
and off-diagonal long-range order. Incorporating short-range
correlations due to interatomic repulsion, this wave function is
shown to allow condensation of zero-point lattice vibrations and
phase rigidity. In the presence of an external velocity field,
such a perfect crystal will develop non-classical rotational
inertia, exhibiting the supersolid behavior. In a sample
calculation we show that the superfluid fraction in this state can
be as large as of order 0.01 in a reasonable range of microscopic
parameters. The relevance to the recent experimental evidence of a
supersolid state by Chan and Kim is discussed.
\end{abstract}


\submitto{Journal of Statistical Mechanics: Theory and Experiment}

\maketitle

\section{Introduction}
Superfluidity (including superconductivity) has remained at the
center of attention of low temperature physics since it was
discovered at the beginning of last century. By now it is well
known that both superfluidity in liquids and gases and
superconductivity in electron systems can all be characterized by
the appearance of off-diagonal long-range order
(ODLRO)\cite{Yang}. In contrast to the diagonal long-range order
(DLRO), ODLRO is a quantum phenomenon with long-range phase
coherence, not describable in classical mechanical terms. On the
other hand, Yang has pointed out the possibility that ODLRO may
occur in a solid, coexisting with DLRO \cite{Yang}. The
possibility for such a supersolid has been explored theoretically
or numerically in 1970's by a number of authors
\cite{Andreev,Chester,Leggett,Saslow,Fernandez}. In particular,
Leggett \cite{Leggett} has suggested to study superfluidity in a
quantum solid in terms of non-classical rotational inertia (NCRI).
These have stimulated efforts to search for superfluidity in solid
helium in the following decades \cite{Meisel}. Recent experiments
of Chan and Kim have shown evidence of NCRI in solid ${}^4{\rm
He}$, either confined in porous medium \cite{ChanNature} or in a
bulk \cite{ChanScience}, with the observed values of the
superfluid fraction (SFF) of the order 0.01. The experimental
progress recently drew much attention to revisiting the theory of
supersolids \cite{Ceperley,Saslow04,Weiss}.

In contrast to the efforts concentrated on defects or vacancies,
in this paper we shall consider a {\it perfect crystal} of
${}^4{\rm He}$, where the number, $N$, of atoms precisely equals
that of the sites, and the single-particle density profile has a
discrete translation symmetry. We shall show that (i) a many-body
wave function can be constructed which indeed exhibits both DLRO
and ODLRO, (ii) in this state short-range correlations due to
interatomic hard-core repulsion lead to a condensation of
zero-point lattice vibrations with long-range phase rigidity,
(iii) in the presence of an external velocity field, this state
shows NCRI, and the associated SFF is estimated to be of order
0.01 in a reasonable range of parameters. Finally we point out our
wave function has features consistent with the experiments
\cite{ChanNature,ChanScience}.

\section{ Many-body wave function for a perfect supersolid\label{wavefunction}} We start
with the construction of a wave function that describes a perfect
supersolid. In a normal crystal individual atoms are localized and
oscillate around their equilibrium positions, which form a
lattice, so that the density profile is periodic, resulting in
DLRO. For helium atoms, their small mass makes the zero-point
oscillations significant. Normally the zero-point motion of
individual atoms is incoherent, as in the Einstein picture of
lattice dynamics. This is what happens in normal solid helium
(under pressure). We propose that at sufficiently low temperature,
due to Bose statistics of ${}^4{\rm He}$ atoms and short-range
correlations arising from interatomic hard-core repulsion, the
zero-point motion of individual atoms may become phase-locked and,
therefore, an ODLRO and phase rigidity will be developed across
the whole system. We shall not address the question of under
precisely what conditions this will happen, but be concentrated on
constructing a wave function that demonstrates that the
coexistence of DLRO and ODLRO is possible in principle.

We use a localized wave function, $\phi({\bf r}-{\bf R}_{i})$, to
describe the zero-point motion of the atom around a lattice point
${\bf R}_i$. Then the coherent zero-point motion of the atoms in
the crystal can be described by the following symmetrized product
of single particle wave functions:
\begin{equation}
{\cal S}\prod\limits_{i=1}^{N}\phi({\bf r}_{i}-{\bf
R}_{i})\label{symmetry},
\end{equation}
with ${\cal S}$ the symmetrization operator with respect to $r_i$.
The symmetrization ${\cal S}$ in Eq. (1), which is absent for an
Einstein crystal, incorporates the fact that all atoms are
identical bosons. The superposition involved in ${\cal S}$ of
permuted products of $\phi({\bf r}_{i}-{\bf R}_{j})$ takes for
granted that the zero-point lattice vibrations of individual atoms
must be {\it coherent}; otherwise it would not make sense to
superpose the permuted terms. Moreover, as in the description of
helium superfluid, we should also include a Jastrow factor to
incorporate short-range pair correlations. Normally the Jastrow
factor is taken to be
\begin{equation}
\prod\limits_{i<j}J(r_{ij})\equiv \prod\limits_{i<j} \exp
\{-\gamma v(r_{ij})\}\label{jastrow},
\end{equation}
with $\gamma>0$ and the exponents proportional to the interatomic
potential $v(r_{ij})$ of the Lennard-Jones type, with a hard-core
repulsion plus a weak attractive part. Usually the product of wave
function (\ref{symmetry}) and (\ref{jastrow}) is used to describe
a perfect quantum solid, namely
\begin{equation}
\Psi={\cal S}\prod\limits_{i=1}^{N}\phi({\bf r}-{\bf
R}_{i})\prod\limits_{i<j}J(r_{ij})\label{Hatree-Fock-Jastrow}.
\end{equation}
Provided that the characteristic width, $a$, of the localized wave
packet $\phi$ is much smaller than the hard-core size $\lambda$,
the wave function (\ref{Hatree-Fock-Jastrow}) can be approximated
by
\begin{eqnarray}
\Psi_0= \frac{1}{{\sqrt N!}} \prod\limits_{i=1}^{N} \left(
\sum\limits_{j=1}^{N}\phi({\bf r}_{i}-{\bf R}_{j})\right)
\prod\limits_{k<l}J(r_{kl})\label{supersolid},
\end{eqnarray}
since the Jastrow factor almost annihilates the cross terms in
which two atoms are in the same localized wave packet at one site.
However, when $a$ becomes comparable to $\lambda$, the wave
function (\ref{supersolid}) incorporates some new features and may
exhibit qualitatively different behavior than the wave function
(\ref{Hatree-Fock-Jastrow}). The above arguments motivate us to
propose ${\Psi_0}$ as our model wave function to describe, at
least approximately, a perfect supersolid, and we will proceed to
show that this wave function allows appreciable Bose-Einstein
condensation in a density periodic state.

\section{ Proof of Bose-Einstein condensation}
From previous experience with ${}^4{\rm He}$ superfluid it is
known that the Jastrow factor can have a Bose-Einstein
condensation into the zero-momentum state
\cite{Retto,Reatto,Reato-Chester,Chester,Fernandez2}. Similarly by
expanding the Jastrow factor in Eq. (\ref{supersolid}):
\begin{equation}
\Psi_0=\prod\limits_{i=1}^{N}\left(\frac{1}{\sqrt{N}}
\sum\limits_{j=1}^{N} \phi({\bf r}_{i}-{\bf R}_{j})\right)
\left(\sqrt{n_{0}}+\cdot\cdot\cdot\right),
\end{equation}
a non-vanishing zero-mode part ($n_0\neq 0$) would give rise to
macroscopic occupation in a single particle state, which has a
periodic density modulation. (The value of the condensate fraction
$n_{0}$ depends on the microscopic details, and in the superfluid
case it was estimated to be of the order $0.01$
\cite{Fernandez2}.) Now let us proceed to prove that the wave
function (\ref{supersolid}), which is of the general form: ($u(
r_{ij})=\gamma v(r_{ij})$)
\begin{equation}
\Psi_{0}\sim \prod\limits_{k=1}^{N}f({\bf
r}_{k})\prod\limits_{i<j}J(r_{ij}) =\prod\limits_{k=1}^{N}f( {\bf
r}_{k}) \prod\limits_{i<j}\exp\{-u( r_{ij})\}, \label{general}
\end{equation}
has a Bose-Einstein condensation on the single-particle state
$f({\bf r})$, provided that $|f({\bf r})|$ nowhere vanishes and
has an upper bound. Assume the same conditions used in
Ref.\cite{Reatto,Reato-Chester}: There exists a positive constant
$\phi$ such that the function $u( r_{ij})$ satisfies $
\sum_{i=1}^{t}u(r_{is})\geq -\phi$, for all $t$, $s$,
$r_{1},r_{2},\cdots$ satisfying $\sum_{i<j\leq
t}u(r_{ij})<\infty$. The proposition is the resulting one-particle
density matrix possesses ODLRO:
\begin{equation}
\label{ODLRO} \lim\limits_{|{\bf r}-{\bf
r}^\prime|\rightarrow\infty}\langle {\bf r}|\rho_{1}|{\bf
r}^\prime\rangle=n_{0}f^{*}({\bf r})f({\bf
r}^\prime)\label{one-particle-matrix},
\end{equation}
with $n_{0}$ finite and positive, where
\begin{equation}
\langle {\bf r}|\rho_{1}|{\bf r}^\prime\rangle=\frac{N}{Q_{N}}\int
\prod\limits_{i=2}^{N}d{\bf r}_{i}\Psi^{*}_{0}({\bf r},{\bf
r}_{2},\cdots) \Psi_{0}({\bf r}^\prime,{\bf r}_{2},\cdots)
\end{equation}
with $Q_{N}$ the normalization constant of $\Psi$. Eq.
(\ref{ODLRO}) implies that the wave function (\ref{general}) has a
Bose condensation (macroscopic occupation) in the single particle
state $f(r)$.

To prove Eq.(\ref{one-particle-matrix}), we notice in the infinite
volume limit,
\begin{equation}
n_{0}=\lim\limits_{V\rightarrow\infty}\frac{1}{V^2}\int
drdr^\prime\frac{\langle {\bf r}|\rho_{1}|{\bf
r}^\prime\rangle}{f^*({\bf r})f({\bf r}^\prime)}
=\frac{N}{V^2}\frac{\zeta_{N+1}}{Q_{N}}, \label{condfrac}
\end{equation}
where $\zeta_{N+1}$ is defined as
\begin{equation}
\zeta_{N+1}= \int \frac{drdr^\prime}{f^{*}({\bf r})f({\bf
r}^\prime)} \prod\limits_{i=2}^{N}d{\bf r}_{i}\Psi^{*}_{0}({\bf
r},{\bf r}_{2},\cdots) \Psi_{0}({\bf r}^\prime,{\bf
r}_{2},\cdots).
\end{equation}
Note that the Jastrow functions in $\zeta_{N+1}$ are given by
\begin{eqnarray}
J(r_{ab})\prod\limits_{k=2}^{N}J(r_{ak})J(r_{bk})
\prod\limits_{2\le i<j} J^2(r_{ij})
\end{eqnarray}
where ${\bf r}_a={\bf r},{\bf r}_b={\bf r}^\prime$. The use of the
inequality $\sum_{i=1}^{t}u(r_{is})\geq -\phi$ allows us to give a
lower bound for $\zeta_{N+1}$:
\begin{eqnarray}
\zeta_{N+1} \geq \frac{e^{-\phi-\Delta}}{\kappa}Q_{N+1},
\end{eqnarray}
where $\Delta=\min u(r)$ and $\kappa=\max |f({\bf r})|^4$.
Therefore the condensate fraction (\ref{condfrac}) has a lower
bound:
\begin{equation}
n_{0}\geq \frac{n^2}{z}\frac{e^{-\phi-\Delta}}{\kappa},
\end{equation}
where the density $n\equiv N/V$ is fixed in the thermodynamic
limit, and $z$ denotes the limit
\begin{equation}
z=\lim\limits_{V,N\rightarrow\infty}\frac{(N+1)Q_{N}}{Q_{N+1}}.
\end{equation}
Note $Q_N$ can be interpreted as the partition function
\begin{equation}
\int\prod\limits_{i=1}^{N}d{\bf
r}_{i}\exp\left\{-\sum\limits_{i<j} u(|{\bf r}_{i}-{\bf
r}_{j}|)+2\sum\limits_{i}\ln|f({\bf r}_{i})|\right\} \nonumber
\end{equation}
of a system of classical particles interacting through the
two-body potential $k_{B}T_{eff}u(r)$ and in an external potential
$2k_{B}T_{eff}\ln|f({\bf r})|$ with fugacity $z$. For our wave
function (\ref{supersolid}), $f(r)\equiv\sum_j \phi({\bf r}-{\bf
R}_j)$ is periodic, positive and everywhere nonvanishing, so the
external potential is also finite everywhere. Hence, the
thermodynamic limit of this classical system exists, and fugacity
$z$ is finite if $n<n_{c}$, where $n_{c}$ is the close-packing
density. In this way we have proved that at $T=0$ our many-body
wave function (\ref{supersolid}) has a finite condensate fraction
$n_{0}$, leading to ODLRO, Eq.(\ref{one-particle-matrix}), a
periodic density profile.

\section{ Non-Classical Rotational Inertia}
Next we study the response of our perfect supersolid to an
external velocity field. Suppose that there are $N$ bosonic atoms
enclosed in a cylindrical annulus with internal radius $R$ and
thickness $h\ll R$. When the cylinder is rotated at a constant
angular velocity $\omega$ about its axis, the free energy
$F(\omega)$ measured in the rest frame is of the form
\cite{Leggett}
\begin{equation}
F(\omega)=F(0)+\frac{1}{2}I_{0}\omega^2-\Delta F(\omega),
\end{equation}
where $F(0)$ is the free energy for $\omega=0$, and $I_{0}=NmR^2$
is the classical rotational inertia. The last term is the NCRI,
which is related to the SFF $\alpha$ via
\begin{equation}
\Delta F(\omega)=\frac{1}{2}I_{0}\omega^2\alpha.
\end{equation}

In the rotating frame (with the azimuthal angles $\varphi_{i} \to
\varphi_{i}+\omega t$), after the gauge transformation
\begin{equation}
\Phi \to \exp\left(-i\sum\limits_{i=1}^{N}\frac{m\omega
R^2\varphi_{i}}{\hbar}\right)\Phi,
\end{equation}
the Schr$\ddot{o}$dinger equation reads \cite{Leggett}
\begin{equation}
i\hbar\frac{\partial}{\partial t}\Phi=\left[\sum\limits_{i}\left(
-\frac{\hbar^2}{2m}\nabla^ 2_{i}+V+\frac{m\omega^2
R^2}{2}\right)+\sum\limits_{ij}U_{ij}\right]\Phi.
\end{equation}
Apart from an additive term, $Nm\omega^2 R/2$, in energy, $\Phi$
satisfies the same equation as $\Phi_0$ in the absence of
rotation, however the boundary conditions are changed to
\begin{equation}
\Phi(\varphi_{i}=0)=\exp\left(-i2\pi\frac{m\omega
R^2}{\hbar}\right)\Phi(\varphi_{i}=2\pi).
\label{boundary-condition}
\end{equation}
One can therefore conclude that the energy levels are periodic
functions of $\omega$. However the free-energy is insensitive to
the twisted boundary conditions, either when it is in the normal
state \cite{Leggett2} or the wave function of all the particles
are localized. Only when (at least) one extended single-particle
state is macroscopically occupied by $N_0$ atoms, the SFF is
finite and given by
\begin{equation}
\alpha=\frac{1}{I_{0}}\frac{\partial^2\Delta F(\omega)}{\partial
\omega^2}.
\end{equation}
This idea is similar to that explaining flux quantization in
superconductors \cite{Yang-flux}.

\begin{figure}[htbp]
\begin{center}
\includegraphics[width=3.0in]
{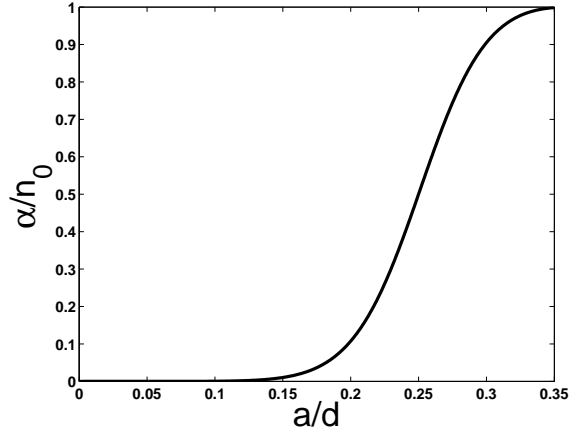} \caption{The ratio of SFF $\alpha$ to
condensate fraction $n_{0}$ as a function of Lindermann's ratio
$a/d$. \label{fraction}}
\end{center}
\end{figure}

\begin{figure}[htbp]
\begin{center}
\includegraphics[width=3.0in]
{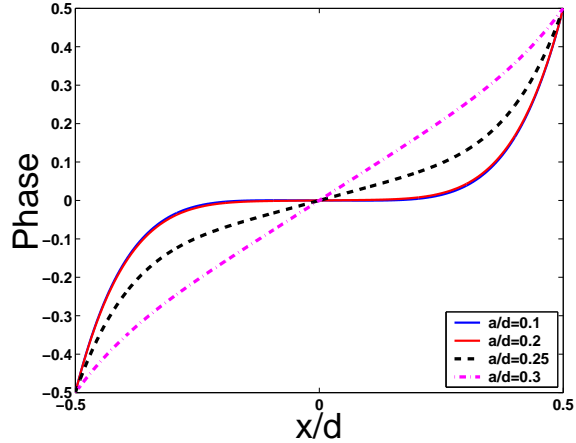} \caption{The phase $\theta$( in units of $c_{0}$) as
a function of $x/d$ in one period for three values of Lindermann's
ratio $a/d$. \label{phase}}
\end{center}
\end{figure}

For the superfluid component, we have
\begin{equation}
\Delta F(\omega)=N_{0}\min\int\limits_{0}^{L}
dx\frac{\hbar^2}{2m}|\nabla\theta|^2\rho,
\end{equation}
where $\sqrt{\rho} e^{i\theta}$ is the order parameter $\langle
x|\rho_1|x \rangle$, and $L$ is the length of the sample. Here as a sample calculation, we only
consider a one dimensional lattice along the circumference
$x$-direction. The one-particle density $\rho$ has a periodic
structure:
\begin{equation}
\rho(x)=\frac{1}{N}\sum\limits_{i=1}^{N} A(x-R_{i}),
\end{equation}
where $|R_{i}-R_{i-1}|=d$, and the function $A$, for simplicity,
is taken to be a localized Gaussian packet:
\begin{equation}
A(x)=\frac{1}{\sqrt{\pi a^2}}\exp \left(-\frac{x^2}{a^2}\right).
\end{equation}
The phase $\theta(x)$ is non-uniform, but satisfies
\begin{equation}
\theta(L)=\theta(0)+\frac{m\omega R L}{\hbar}.
\end{equation}
Write $\theta=\theta_{0}+\delta\theta$, where $\theta_{0}=c_0x$
with $c_{0}=m\omega R/\hbar$. It is natural to assume that
$\delta\theta$ has a periodic dependence on $x$ with period $d$.
To minimize $\Delta F(\omega)$, the phase gradient becomes large
only in regions with low density. Therefore the SFF is expected to
be less than the condensate fraction and to increase with
decreasing density modulation. We apply a simple variational
method to prove this. By fixing a global phase we set
\begin{equation}
\delta\theta\left(\frac{d}{2}\right)
=\delta\theta\left(-\frac{d}{2}\right)=\delta\theta
\left(\frac{(2k+1)d}{2}\right)=0,
\end{equation}
For $\delta\theta$ in the interval $[-d/2,d/2]$, we take
\begin{equation}
\delta\theta=c_{0} d \left[c_{1}\frac{x}{d}
+c_{3}\left(\frac{x}{d}\right)^3
+c_{5}\left(\frac{x}{d}\right)^5\right]
\end{equation}
as a trial function, in which the parameters should satisfy the
constraint that $c_{5}=-16c_{1}-4c_{3}$. Noting that $N_{0}\hbar^2
c_{0}^2/m =n_{0}I_{0}\omega^2$, the SFF $\alpha$ can be expressed
as
\begin{eqnarray}
\frac{\alpha}{n_{0}}= N
\min\limits_{c_{1},c_{3}}\int\limits_{-d/2}^{d/2}dx\left
|(1+c_{1})+3c_{3}\left(\frac{x}{d}\right)^2
+5c_{5}\left(\frac{x}{d}\right)^4\right|^2\rho. \label{minimized}
\end{eqnarray}
The right-hand side of Eq.(\ref{minimized}) is quadratic in
$c_{1}$ and $c_{3}$ and can be easily minimized. The resulting SFF
is a function of the Lindermann radio $a/d$. We plot it in Fig.
\ref{fraction}. The phase $\theta$ as a function of $x$ is shown
in Fig. \ref{phase} for three typical values of $a/d$. Our results
show that the SFF is vanishingly small when $a/d<0.1$. When
$a/d>0.1$, the SFF experiences a rapid increase and reaches the
condensate fraction $n_{0}$ when $a/d$ is around $0.35$. Our
result, obtained analytically by a simple variational calculation,
is consistent with the previous numerical result obtained for
three dimensional fcc lattice \cite{Saslow,Saslow04}.

\section{Summary}
In summary, we have suggested that in a perfect quantum bosonic
solid, under favorable conditions, a conspiracy of Bose
statistics, significant zero-point fluctuations and short-range
correlation due to interatomic hard-core repulsion may lead to
ODLRO and macroscopic phase coherence, coexisting with crystalline
DLRO and leading to the supersolid behavior, e.g. the appearance
of NCRI under vessel rotation. We have proposed a many-body wave
function and proved that indeed both DLRO and ODLRO coexist in
such a state. The interatomic potential of the Lennard-Jones type,
having a large hard core and being mostly attractive outside, with
favorable parameters is arguably credential to choose our state
before other non-supersolid states. We note that with this
mechanism for supersolid the more closely packed the lattice is,
the larger the SFF $\alpha$ is. Also the presence of a small
number of vacancies or impurities (e.g. ${}^3{\rm He}$) is harmful
to condensation of zero-point vibrations, and thus reduces the
SFF. Observationally, these two features may be exploited as
qualitative indication of the underlying mechanism suggested here
for supersolid behavior.

The mechanism for supersolid proposed in this note (section
\ref{wavefunction}) can be understood intuitively in the language
of path integral. Consider, say, a one dimensional lattice of
spacing $d$. If $d-2a<\lambda$, where $a$ is the amplitude of
zero-point vibration, then the probability amplitude for two
nearby atoms oscillating completely out of phase will become
negligible due to huge potential energy of the hard-core
repulsion. Therefore, it is the phase-locked trajectories of
oscillating atoms that have lower potential energy and dominate
the path integral, namely only nearly in-phase zero-point motion
of all atoms in the crystal are favorable. It should be noticed
that these phase-locked trajectories are quantum fluctuating
pathes describing the ground state in the framework of path
integral, which should be distinguished from the acoustic phonon
excitation. It is these phase-locked trajectories that give rise
to long-range phase coherence or phase rigidity as described by
the first factor in Eq.(\ref{supersolid}). We call this phenomenon
as {\it condensation of zero-point lattice vibrations}.

In recent experiments of Chan and Kim
\cite{ChanNature,ChanScience} that show NCRI of solid ${}^4{\rm
He}$, their sample is claimed to be ultrahighly pure (with a
stated ${}^3{\rm He}$ impurity of 0.3 parts of per million). Also
parameterwise we note \cite{Glyde} that in solid ${}^4{\rm He}$
the hard core radius is a big fraction ($> 0.65$) of the lattice
spacing; and the ratio $a/d$ can be as big as $0.22$ in a variety
of situations. As mentioned above, all these features are
favorable to our suggested mechanism and to get an appreciable
SFF. It is arguable that the condensation fraction $n_0$ is of the
order 0.01 as estimated in Ref. \cite{Fernandez2}; then in
accordance with our estimation, Fig. \ref{fraction}, the SFF of
our state is appreciable and may reach the order of 0.01 if $a/d$
is not less than 0.25. (A likely explanation of why Ref.
\cite{Leggett} found the SFF $\alpha \leq 10^{-4}$ is that the
estimate used an analogy of the exchange effect for ${}^3{\rm He}$
in a normal rather than supersolid phase.) Since the SFF depends
sensitively on the microscopic parameters, to measure the lattice
constant and the NCRI at the same time would be very desirable.

The authors would like to thank Professor C. N. Yang for
encouragement and helpful discussions. HZ was in part supported by
NSFC (No. 10247002 and 10404015) and YSW by the Spring Sunshine
Visiting Professorship.

\end{document}